# Scaling of the supercooled dynamics and its relation to the pressure dependences of the dynamic crossover and the fragility of glass-formers

R. Casalini[1,2] and C.M. Roland[2]

[1]George Mason University, Chemistry Department, Fairfax VA  22030
[2]Naval Research Laboratory, Chemistry Division, Code 6120, Washington DC  20375-5342

**Abstract**

Master curves of the relaxation time, $\tau$, or viscosity, $\eta$, versus $T^{-1}V^{-\gamma}$, where $T$ is temperature, $V$ the specific volume, and $\gamma$ a material constant, are used to deduce the effect of pressure on the dynamic crossover and the fragility. The crossover is determined from the change in slope of derivative plots of the relaxation times or viscosities. We confirm our previous findings that the value of $\tau$ or $\eta$ at the crossover is independent of both $T$ and $P$; that is, the dynamic crossover is associated with a characteristic value of the relaxation time. Previous determinations were limited to liquids having crossovers occurring at large values of $\tau$ ( $> 10^{-6}$ s), whereas by interpolating within $T^{-1}V^{-\gamma}$ space, we extend the analysis to smaller values of the crossover time. Using the superpositioned data, the dynamic crossover can be observed in isochoric data, where it is found that the relaxation time at the crossover for constant volume is equivalent to the value obtained under (the more usual) condition of constant pressure.

Similarly, from the scaling analysis, isobaric relaxation times at high pressure are deduced from experimental measurements at atmospheric pressure. We find for all glass-formers studied, that the fragility (normalized temperature dependence of $\tau$ or $\eta$) is a decreasing function of pressure. This conclusion is less subject to uncertainties in the measurements than published determinations of the pressure coefficient of fragility. Finally, we show that an empirical function, having the form of the Cohen-Grest relation but without connection to any free volume model, parameterizes the master curves, and accurately describes the data over all measured conditions.

**Introduction**

The topic of the glass transition has been at the center of discussions in condensed matter physics for many years. A still open question is whether the sudden, yet progressive, slowing down of the dynamical properties (viscosity, relaxation time, etc.) is due to an underlying thermodynamical transition. The glass transition temperature, $T_g$, is usually defined empirically, with its value depending on the thermal history of the material (cooling rate, aging, etc.), suggestive of its kinetic nature. Effectively, the glass transition corresponds to the condition whereby the material timescale becomes longer than the duration of a typical experiment. However, most theories and models consider the actual transition to occur at a different temperature, one lying either well below $T_g$, (and therefore not directly observable) or at some higher temperature, $T_B$ (other subscripts may apply depending on the model). Signatures for $T_B$ are generally difficult to observe since there is no clear discontinuity in physical properties, as might be expected for a thermodynamic transition. One manifestation of $T_B$ is the "dynamic crossover", which is discussed below.

A physical description of supercooled liquids currently in vogue is based on the energy landscape. In this framework, cooling of a liquid causes progressive entrapment in potential wells, so that the slowing down of the dynamics is related to a progressive decrease of available configurations. Since the energy landscape is related to intermolecular distances, in a typical isobaric measurement, the height of the potential energy barriers will change with density in a manner dependent on the "nature" of the potential. Therefore, for the isobaric temperature dependence of dynamical properties, two convoluted effects determine the number of accessible configurations: changes in thermal energy which determine the accessible region of the energy landscape, and changes of the energy landscape itself arising due to the variation of density with temperature. These convoluted effects are especially evident in materials exhibiting a strong dependence of their relaxation time on density at fixed T. This is indicative of a strong dependence of the potential energy on intermolecular distances.

A popular metric used to classify glass formers is fragility, *m*, usually defined as [1]

$$m = \frac{d\log(x)}{d\left(T_g/T\right)}\bigg|_{T_g} \tag{1}$$

where $x$ represents the relaxation time ($\tau$), fluidity ($\eta^{-1}$) or other dynamical quantity. Per common usage, strong glass-formers are those having a weak dependence (small $m$), while for fragile liquids $x$ has a strong dependence on $T_g/T$ (large $m$). The isochoric (constant volume) fragility, $m_V$, is directly connected to the "shape" of the energy landscape, although this is not necessarily the case for the isobaric fragility, $m_P$, wherein temperature and density effects are convoluted. Herein, we elaborate on this point, by assessing differences between the two fragilities for various materials, and examining how the convolution of temperature and density effects accounts for the observed pressure dependences of the isobaric fragility.

An interesting phenomenon observed at temperatures above $T_g$ is the dynamic crossover. When measured over many decades of frequency, a clear change is evident in the temperature dependence of various properties, including $\tau$ and $\eta$ [2] A determination of the crossover temperature is obtained using the model-independent derivative function introduced by Stickel et al. [3], which for the relaxation time is

$$\phi_T = \left[\frac{d\log(\tau)}{d(1000/T)}\right]^{-\frac{1}{2}} \tag{2}$$

Various theoretical models anticipate, or at least interpret, the dynamic crossover: (i) the liquid-liquid transition postulated for polymers by Boyer[4] (although see [5]); (ii) the crossover from free diffusion to landscape dominated diffusion at $\tau \sim 10^{-9}$s, as predicted by the energy landscape model, first proposed by Goldstein[6]; (iii) the percolation of "liquid-like cells", according to the Cohen-Grest free-volume model[7,8]; (iv) a marked increase in the degree of intermolecular cooperativity, according to the Coupling Model of Ngai [9, 10]; and (vi) the divergence of the viscosity according to Mode Coupling Theory (MCT).[11] (Note this divergence may not actually be observed due to a transition to hopping dynamics [12]).

Recent high pressure measurements led to the discovery that the value of both the dielectric relaxation time [13] and the viscosity [14] at which the crossover occurs are independent of *P* and *T*. Using a modified Stickel function,

$$\phi_P = \left[\frac{d\log(\tau)}{dP}\right]^{-\frac{1}{2}} \tag{3}$$

a clear change of behavior is evident at a fixed value of $\tau$ or $\eta$.

More recently [15,16] we found that $\log(\tau)$ data for various T and P yield a master curve when plotted versus the parameter $T^{-1}V^{-\gamma}$, where $V$ is the specific volume and $\gamma$ is a material-specific parameter. From an analysis of 18 glass-forming liquids and polymers, we find $0.14 \leq \gamma \leq 8.5$. These results are important both from a theoretical point of view, by offering a connection to the nature of the intermolecular potential, [15, 17, 18, 19, 20], and for practical reasons, because the superpositioned data allow predictions for conditions for which experiments may be difficult. For example, high-pressure dielectric data do not extend above $10^7$ Hz, due to experimental limitations. And measurements are either isothermal or isobaric, with isochoric data having to be constructed by interpolation. However, using the general result that $\log(\tau) = \Im(TV^\gamma)$, the properties over a wider pressure range can be obtained, without extrapolating beyond the measured $\tau$, the widest range of the latter invariably being for ambient pressure.

In the following, we use the $\log(\tau) = \Im(TV^\gamma)$ relation to reanalyze the *T*- and *V*-dependences for glass formers which have been measured under high pressure. In this manner, we can explore the consequences regarding the dynamic crossover and fragility. In particular, starting from atmospheric pressure measurements at varying temperature, we predict the isobaric behavior at any pressure $\hat{P}$, by finding numerically the *T* such that $TV^\gamma(0.1 \text{ MPa}) = TV^\gamma(\hat{P})$. In the analyses, the *V*(*T*,*P*) behavior is described using the Tait equation of state. Similarly, the isochoric behavior at a volume $\overline{V}$ is predicted by calculating for each value of $\log(\tau)$ the *T* conforming to the condition $TV^\gamma(0.1 \text{ MPa}) = T^{-1}\overline{V}^{-\gamma}$.

**Results and discussion**

*Dynamic crossover*

*Phenolphthalein-dimethyl-ether (PDE).* The dielectric relaxation time for PDE was measured over a broad range of temperature and pressure, at atmospheric pressure by varying temperature [21] and isothermally with varying pressure [13]. Together with the experimental determination of *V(T,P)*, we were able to show that the superpositioning condition is valid for PDE with γ=4.5; i.e., $\log(\tau) = \Im(TV^{4.5})$.[15] Using this result, we calculate (Fig.1) the respective temperature behaviors of τ for isochoric (solid lines) and isobaric conditions (dotted lines), using the P = 0.1 MPa data (points). From these calculations, it is also possible to predict the behavior at negative pressure, as shown for the isochoric curves toward the right of the isobar at 0.1 MPa. Although negative pressure is a legitimate thermodynamic condition for condensed matter, in practice obtaining data for P < 0 is difficult due to cavitation and others experimental limitations [22]. Note that none of the results in Fig. 1 extend beyond the range of τ which were actually measured.

Previously, we showed by calculating the function $\phi_P$ (eq.(3)) for isothermal measurements on PDE at elevated *P*, that a dynamic change occurs at the same $\tau_B$ as the crossover determined from $\phi_T$ (eq.(2)) for atmospheric pressure. Here, using the $\log(\tau) = \Im(TV^\gamma)$ relation, we extend the high pressure measurements to a more extended frequency range, in order to verify this finding by calculating the function $\phi_T$ for high pressure isobars. In Fig.2, the function $\phi_P$ is displayed for isobars both at atmospheric pressure and as calculated at 100 and 400 MPa; the temperature range of each is given in Table 1. All curves reveal the presence of a dynamic crossover. To obtain a consistent determination of $\tau_B$, we fit lines to the high and low *T* ranges (solid lines), with $T_B$ taken as their intersection (indicated by the arrow). The fitting is carried out over the same range of frequencies for each isobar. (Note that this estimate of $T_B$ yields slightly smaller values of $\tau_B$ than the somewhat different methods used in other publications). From the values of $\tau_B$(Table1), it is evident that the present results confirm our previous finding that $\tau_B(T,P)$ is essentially constant for a given material.[13]

Also included in Fig.2 is the $\phi_P$ calculated for the isochoric curve at $V$=0.7286 cm$^3$/g (the specific volume at the ambient pressure $T_g$). Here again a dynamic crossover is observed; moreover, and somewhat unexpectedly, we find that $\tau_B$ is equivalent to its value for the isobaric curves. The existence of the crossover under constant volume conditions, together with its existence at constant temperature, confirms the findings that this phenomenon reflects the interplay of both thermodynamic variables.

*Chlorinated biphenyl (PCB62).* Dielectric relaxation times for PCB62 (a chlorinated biphenyl having 62% chlorine by weight) at various temperatures and both atmospheric and elevated pressures have been presented previously [23], and more recently, we measured the $V(T,P)$ dependence [24]. From the analysis of the PCB62 data, using either the Adam-Gibbs model [25,26,27] or the model independent function $\phi_P$ [13], we determined that $\tau_B$ was independent of both $T$ and P. From the $\tau(T,P)$ and $V(T,P)$ data, the scaling $\log(\tau) = \Im(TV^\gamma)$ was verified, with $\gamma$=8.5.[24] Using the scaling properties, in the manner for PDE, we calculate isobars at higher pressure, as well as one isochore (V=0.6131 ml/g). These are shown in Fig.3 (upper panel), along with the corresponding Stickel function (lower panel). The dynamic crossover is evident in each curve, and using the method described above for PDE, the crossover temperatures and relaxation times $\tau_B$ were obtained (Table1). For all combinations of T and P, as well as for the isochoric condition, $\tau_B$ is essentially constant, $\log(\tau_B$ /s)=-5.9±0.5. This is somewhat less than the value estimated previously [13], due only to the different methods used to determine $T_B$.

*Propylene carbonate (PC).* Dielectric relaxation of PC has been measured at atmospheric pressure [21,28], and more recently at high pressure.[29] From the latter, $\tau$ and the ionic conductivity ($\sigma$) were found to exhibit a crossover at a characteristic value that is pressure-independent. From superposition of the relaxation times, $\gamma$ = 3.7. From this, we calculate $\tau$ for high pressure over an extended frequency range, in the manner done for PDE and PCB62. Three isobars at $P$ = 0.3, 0.6 and 1 GPa are displayed in Fig.4 (upper panel), together with an isochoric curve (V=0.7562 cm$^3$/g). The calculated $\phi_T$ are shown in the lower panel. $T_B$ was determined using the same procedure described above, the results are listed with the corresponding values of log($\tau_B$) in Table 1. Again, the relaxation time associated with the change in dynamics is sensibly independent of *T* and *P*, log($\tau_B$) =-7.0±0.3. An equivalent value is obtained for the isochoric curve.

In conclusion, from analysis using the relation $\log(\tau) = \Im(TV^\gamma)$ for PDE, PCB62 and PC, we confirm our previous finding that $\tau_B(T,P) \approx$ constant, with the obtained values equal to the $\tau_B$ directly measured. Thus, we can use this method to analyze other glass-formers, in which the range of the available experimental data is insufficient to probe the dynamic crossover under high pressure.

*Cresolphthalein-dimethyl-ether (KDE)*. Dielectric relaxation times for KDE have been measured.[21,30] The material has a chemical structure very similar to that of PDE, but $\tau_B$ at atmospheric pressure is more than two decades shorter. As a result, the function $\phi_P$ for KDE [31] shows no indication of a change of behavior in isothermal data measured at frequencies up to 10$^5$ Hz. However, we can take advantage of the master curve of $\log(\tau)$ vs. $TV^\gamma$, with γ =4.5 for KDE, to determine $\tau_B$ at elevated pressures. In the upper panel of Fig.5 are shown three isobars (experimental data for ambient pressure, along with two calculated curves) and an isochoric curve. In the lower panel is the function $\phi_T$ calculated for each curve. For all conditions, there is a crossover occurring at log($\tau_B$)=-6.4±0.3. The particular values of $T_B$ and log($\tau_B$) for each curve are listed in Table 1. These results are consistent with the absence of a crossover in high pressure measurements done at lower frequencies.

*1,1'-di(4-methoxy-5-methylphenyl)cyclohexane (BMMPC)*. BMMPC has the largest value of the scaling exponent, γ =8.5. Since a dynamic crossover in BMMPC is well-established [8], it is interesting to evaluate the *P*- and *T*-dependences of $\tau_B$. A large value of γ indicates a more dominant role for volume, so we expect stronger *T*- and *P*-dependences for $\tau_B$ than for the other materials. Three isobars and an isochoric curve are shown in the upper panel of Fig.6, with the corresponding $\phi_T$ in the lower panel. Following the analysis described above, $\tau_B$ for BMMPC is found to be invariant to both *T* and *P*, log($\tau_B$) = -6.1±0.5. This invariance of $\tau_B$ is maintained even under constant volume conditions; thus, notwithstanding the prominent role of volume in the dynamics of BMMPC, the characteristic relaxation time at the crossover is unchanged when $V$ is maintained constant.

*Salol*. Salol is one of the few materials for which high pressure data are available for different experimental properties such as viscosity [32] photon correlation spectroscopy (PCS)[33] and dielectric relaxation [34] We have recently shown from dielectric relaxation measurements that $\gamma=5.2$, so it is of interest to test this scaling relation for the other experimental quantities. In Fig.7 $\log(\tau)$ from PCS [33] (lower panel) and $\log(\eta)$ [32,35] (upper panel) are plotted versus $T^{-1}V^{-5.2}$. It is evident that the PCS data superpose accurately using the same $\gamma$ parameter determined from dielectric relaxation. However, for the viscosity there are some deviations. While the data converge at low viscosities, a different behavior is observed for higher $\eta$. In particular, the two isotherms at the highest $T$ seem to superpose, but they differ from the isobar at atmospheric pressure and the two lower temperature isotherms. Clearly, there is no value of $\gamma$ which will yield a single master curve of the $\eta$ data. The reason for this is unknown, although previously we had noted an apparent decoupling of $\tau$ and $\eta$ at high pressure.[34] Given that the dielectric and photo-correlation data both superimpose for the same value of $\gamma$, it seems that further measurements are required to determine whether the breakdown of the $\gamma$–scaling for $\eta$ is real. It is interesting that the $\log(\tau)$ and $\log(\eta)$ are both linear versus $T^{-1}V^{-5.2}$, for $\tau$ longer than $\tau_B$ (or $\eta>\eta_B$). This would be consistent with thermally-activated (Arrhenius) behavior with and activation energy, $E_a \propto V^{5.2}$. Similar behavior has been observed for other materials. [15]

Calculating $\phi_P$ for the isothermal $\eta$ data, a dynamic crossover is evident at the same viscosity as for the atmospheric pressure data. However, since the $\eta$ in Fig.7 do not entirely superpose, this result cannot be confirmed for higher pressures. We can carry out the analysis for the dielectric results. In the upper panel of Fig. 8 is the isobar at atmospheric pressure [3], together with two isobars and an isochoric curve, all calculated using $\log(\tau)$ vs. $TV^{5.2}$. The $\phi_T$ calculated for each curve is shown in the lower panel, with the crossover temperature, listed in Table 1, determined from the intersection of the linear fits. From this analysis, we find $\log(\tau_B)$=-6.4±0.3 for all P, as well as for the isochoric condition.

$T_B(P)$ extracted from the dielectric data are compared to the $T_B(P)$ determined from the viscosity [14] in the inset to Fig. 8. The values are very close, confirming our previous finding, as well as the use of the $\log(\tau) = \Im(TV^\gamma)$ scaling for the prediction of $\tau(T,P)$. This agreement, despite the lack of superpositioning of the viscosities, likely reflects the fact that the deviation from $\log(\eta) = \Im(TV^\gamma)$ prevails only for $\eta > \eta_B$ (Fig.7).

*-Isobaric and isochoric fragility*

We can take advantage of the scaling of the relaxation times to analyze the fragility, in particular its pressure dependence and any differences for isochoric and isobaric conditions. Generally, it has been found for most of materials that $m_P$ either decreases or is constant as pressure increases. Usually, $m_P$ is calculated from the activation volume, $\Delta V^\#$ ($= RT \, \partial \ln(\tau)/\partial P$), according to

$$m_P = \frac{\Delta V^\#}{\frac{dT_g}{dP} R \ln(10)} \tag{4}$$

The use of this equation is subject to larger errors than a direct comparison of isobars, the latter possible through use of the scaling relation, $\log(\tau) = \Im(TV^\gamma)$. To avoid extrapolating, we calculate fragilities at temperatures $T_\alpha$ such that $\tau(T_\alpha)=10$s. This yields a smaller value of $m$ than using $\tau(T_g)=10^2$s, but it does not affect the $P$- or $V$-dependences.

In Fig.9 are shown $\tau$ (measured at atmospheric pressure and calculated for $P>0.1$ MPa ) for salol, PC, BMMPC, PDE, KDE and PCB62, as a function of the inverse temperature normalized by $T_\alpha$. For all six materials, the fragility decreases with increasing pressure. This is shown more clearly in Fig.10, where $m_P$ is plotted versus $P$. Fig.9 also includes isochoric curves, calculated for the V at which $\tau=10$s at $P=0.1$ MPa. The values for $m_V$, listed in Fig.10, are all smaller than the corresponding isobaric $m_P$. Moreover, there is a rough positive correlation between $m_P$ and $m_V$.

In light of our finding that the isobaric fragility always decreases with pressure, we reconsider 1,2-polybutadiene (PB) [36], the sole case of an $m_P$ (determined using eq.(4)) reportedly increasing with $P$. The $\tau$ measured for PB for four isotherms at varying $P$ are displayed in Fig.11 [36], along with the best fit (solid line) of the data to the empirical function [37]

$$\tau = \tau_0 \exp\left(\frac{DP}{P-P_0}\right) \quad (5)$$

where $\tau_0$, $D$ and $P_0$ are constants. This fitting equation was employed to estimate the activation volume and the value of the pressure at which $\tau=1$s [36]. Using the relation $\log(\tau) = \Im(TV^{1.9})$ [15], we calculate isotherms for elevated pressures (shown as solid symbols in Fig.11). There is a significant difference between the values of $P(\tau=1s)$ determined from the master curve of $\log(\tau)$ versus $T^{-1}V^{-1.9}$, and those calculated via eq.(5). This is especially the case at lower temperature, where the data are scarce. These differences lead to large differences in the pressure coefficient of $T_g$ (10% and 15% larger than $dT_g/dP$ determined using eq.(5)). From the calculated isobars, $m_P$ is found to decrease with P (see insert), which agrees generally with experimental data on polymers [38] (although contrary to the results reported in [36]). Thus, with the exception of the strongly H-bonded glycerol [39], fragility always decreases with pressure

The expressions for the isobaric and isochoric fragilities can be rewritten as

$$m_P = \left.\frac{\partial \log(\tau)}{\partial (Y/Y_g)}\right|_{Y_g} (1+\gamma\alpha_P T_g) \quad (6)$$

$$m_V = \left.\frac{\partial \log(\tau)}{\partial (Y/Y_g)}\right|_{Y_g} \quad (7)$$

where $\left.\frac{\partial \log(\tau)}{\partial (Y/Y_g)}\right|_{Y_g}$ defines the slope in a normalized master curve (fragility plot with $Y$ ($=T^{-1}V^{-\gamma}$) replacing T) [15]). It follows from the superposition relation ($\log\tau = \Im(TV^\gamma)$), that the isochoric fragility $m_V$ is a constant. Since γ is a constant, the pressure-dependence of $m_P$ is governed by the product of the isobaric thermal expansion coefficient and the glass temperature, and for smaller γ, this P-dependence will be smaller. Eq.(6) can be also be rewritten in terms of the ratio of the isochoric ($E_V = R \frac{\partial \ln(\tau)}{\partial T^{-1}}\bigg|_V$) and isobaric ($E_V = R \frac{\partial \ln(\tau)}{\partial T^{-1}}\bigg|_P$) activation enthalpies. Using the relation [15]

$$(1+\gamma \alpha_P T_g) = \left(\frac{E_V}{E_P}\bigg|_{T_g}\right)^{-1} \tag{8}$$

equation (6) becomes

$$m_P = \frac{\partial \log(\tau)}{\partial (Y/Y_g)}\bigg|_{Y_g} \left(\frac{E_V}{E_P}\bigg|_{T_g}\right)^{-1} \tag{9}$$

From eq. 9, a decrease of $m_P$ with increasing pressure implies that of the ratio $E_V/E_P\big|_{T_g}$ is an increasing function of *P*. This means that at higher pressure, the relative influence of *T*, in comparison to *V*, becomes magnified. Such an inference seems to be in accord with experimental observations [20].

Recently, De Michele et al. [18] carried out simulations of a binary mixture of soft spheres interacting with an inter-particle repulsive potential, ~$r^{3\gamma}$. They calculated diffusion constants for different degrees of "softness" of this potential (γ= 6, 8, 12, 18), and found that the isochoric fragility is independent of γ. This lack of correlation is consistent with our results herein. For example, PDE and KDE are associated with the same exponent γ, but have different fragilities. No more than a rough correlation between $m_V$ and γ can be expected, since other factors affect the fragility. The change of fragility can be ascribed to a change in the number of available configurations. In particular, Speedy [40] derived a direct proportionality between the two quantities. Sastry [41], on the other hand, found that the fragility was proportional to the square of the number of available configurations. A discussion of this subject can be found in [42].

Comparing the isochoric and isobaric behaviors of different materials at the same *T*, the isochoric τ is larger because of fewer available configurations, and it will have a weaker *T*-dependence, since changes in the available configurations with *V* are omitted. That is, the number of configurations changes only due to the possibility of exploring different parts of the energy landscape, as governed by the available thermal energy. Similarly, we note that at higher pressure, variations in τ due to the volume changes accompanying changes in T are reduced, and thus a corresponding contribution to the isobaric fragility is removed. In the limit of very high pressure, two possible scenarios are feasible: (i) $T_g \to T_g(\infty)$ and $\alpha_P \to 0$, from which it follows that $m_P \to m_V$; or (ii) $\alpha_P T_g \to \alpha_P T_g(\infty)$=const, and $m_P \to m_V(1+\gamma\alpha_P T_g(\infty))$. The second case is supported by the fact that for many materials, at atmospheric pressure $\alpha_P T_g \sim 0.16 - 0.19$ [43]. This also accounts for the correlation we observe between $m_P$ and $m_V$ at atmospheric pressure. The relation between fragility and the available configurations is consistent with the independence of $m_V$ on *V*. A recent calculation by Speedy [44], who assumed a form for the intermolecular potential similar to that of De Michele et al.[18], indicated that the number of available minima in the energy landscape is indeed independent of V.

*-Phenomenogical descriptions of* $\log[\tau(Y)]$

Since the τ( *T*, *V*) behavior can be expressed as a single function of $TV^\gamma$, it is of interest to find an analytical form for this function. An obvious starting point is to consider equations which provide a good description of $\log[\tau(T)]$; that is, in the limit γ→0. The most common function is the Vogel-Fulcher equation [45,46]; however, it is well know that a single Vogel-Fulcher equation cannot account for the temperature behavior over the entire supercooled regime, from $T_g$ to above $T_B$.

An alternative expression is the Cohen-Grest (CG) equation [7, 47]. It is one of the few equations that can describe τ(T) over a broad range, using only four adjustable parameters [8,48]. Substituting in $Y^{-1}$ for *T*, it can be written

$$\log(\tau) = A + \frac{B}{Y^{-1} - Y_0^{-1} + \left[\left(Y^{-1} - Y_0^{-1}\right)^2 + CY^{-1}\right]^{\frac{1}{2}}} \qquad (10)$$

where $A$, $B$, $C$ and $Y_0^{-1}$ are constants.

The best fits of eq.(10) to the relaxation times of the six liquids considered herein (showing only the atmospheric pressure data for simplicity) are displayed in Fig.12, with the CG parameters listed in Table 2. The residuals are plotted in the insets. It can be seen that the deviation of the fitted curves are less than $\pm 0.1$ over the entire range.

Although the CG equation was derived from a free-volume model of the glass transition [7,47], eq.(10) herein is an empirical modification, intended only to parameterize the $\tau(TV^\gamma)$ data. $T_0$ in the CG equation has been found to correspond to the crossover temperature, $T_B$.[8] Whether the form of eq.(10) can be linked to any theoretical model remains to be seen.

**Conclusions**

We have shown that master curves of the relaxation time or viscosity, having the form $\log(\tau) = \Im(TV^\gamma)$ can be exploited to analyze the dynamics in supercooled liquids over broad ranges of T, P and V. A functional form for the superpositioning variable is obtained by empirical modification of the CG equation (eq.(10)), without implying any connection to the free volume concepts underlying the original CG model. The focus herein is on the high frequency crossover and the dependence of the fragility on P and V. Regarding the crossover: (i) We have confirmed previous findings for PDE, PCB62, PC and salol that the crossover transpires at a fixed value of the relaxation time, $\tau_B$ (or the viscosity, $\eta_B$), independent of the particular conditions of T and P; (ii) We have found that the dynamic crossover is evident in isochoric curves.

From analysis of temperature dependences, we find: (i) The isobaric fragility decreases with increasing pressure; this seems to be a general result for van der Waals glass-formers. (ii) The isochoric fragility is independent of pressure. (iii) The isochoric fragility is significantly less than the isobaric fragility. (iv) There is a rough correlation between the difference of between $m_P$ and $m_V$ and the magnitude of the scaling exponent γ. This correlation can be accounted for from eqs.(6) and (7), and considering that $α_P T_g$~0.16-0.19 [43]. (v) The decrease of the fragility with pressure is related to the magnitude of the increase with P in the ratio of the isobaric and isochoric activation enthalpies. These results are consistent with the idea that the fragility (and therefore the slowing down of the dynamics) is related mainly to the number of available configurations, rather than to the shape of the intermolecular potential. The latter is only one of the factors governing the configurational entropy.

**Figure Captions**

**Fig.1** Dielectric relaxation time for PDE. Solid lines are calculated isochoric curves for $V_1$=0.72865, $V_2$=0.7380, $V_3$=0.7602 and $V_4$=0.7795 cm$^3$/g. Dotted lines are calculated isobaric curves for P= 100 and 400 MPa. Points (■) are for atmospheric pressure.

**Fig.2** Stickel function, $\phi_T$, for dielectric relaxation time of PDE. The lines indicates the respective low and high T linear fits, done over the range -2.27 < $\log_{10}(\tau[s])$ < 1.68 and -8.76 < $\log_{10}(\tau[\sigma])$ < -5.08, whose intersection defines the dynamic crossover (values of $T_B$ and $\log(\tau_B)$ listed in Table 1).

**Fig.3** Upper panel: dielectric relaxation times for PCB62 (experimental data for 0.1MPa; other isobars and isochoric curve at V=0.6131ml/g were calculated). Dotted line indicates the average of $\log(\tau_B)$=-5.9 for the different curves. Lower panel: Stickel function, with low and high T linear fits, done over the range -5.42 < $\log_{10}(\tau[s])$ < 2.169 and -8.87 < $\log_{10}(\tau[s])$ < -5.98, respectively. The vertical dotted lines in both panels represent the dynamic crossover (see Table 1).

**Fig.4** Upper panel: dielectric relaxation time for PC (experimental data for 0.1MPa; other isobars and isochoric curve were calculated). Dotted line indicates the average of $\log(\tau_B)$ for the different curves. Lower panel: Stickel function, with low and high T linear fits, done over the range -6.14 < $\log_{10}(\tau[s])$ < 0.63 and -10.21 < $\log_{10}(\tau[s])$ < -8.03, respectively. The vertical dotted lines in both panels represent the dynamic crossover (see Table 1).

**Fig.5** Upper panel: dielectric relaxation time for KDE (experimental data for 0.1MPa; other isobars and the isochoric curve at V=0.7709 ml/g were calculated). Dotted line indicates the average of $\log(\tau_B)$=-6.35 for the different curves. Lower panel: Stickel function, with low and high T linear fits, done over the range -4.62 < $\log_{10}(\tau[s])$ < 2.72 and -9.4 < $\log_{10}(\tau[s])$ < -7.28, respectively. Vertical dotted lines indicate the dynamic crossover (Table 1).

**Fig.6** Upper panel: dielectric relaxation time for BMMPC (experimental data for 0.1MPa, other isobars (200 and 600 MPa) and the isochore at V=0.9032 ml/g were calculated. Dotted line indicates the average of $\log(\tau_B)$=-6.1 for the different curves. Lower panel: Stickel function, with low and high T linear fits, done over the range -4.68 < $\log_{10}(\tau[s])$ < 3.85 and -8.55 < $\log_{10}(\tau[s])$ < -6.4, respectively. Vertical dotted lines indicate the dynamic crossover (Table 1).

**Fig.7** Upper panel: Scaled viscosity data for salol measured isothermally at high pressure [32] and at atmospheric pressure with varying temperature [35]. The dotted line corresponds to the value of $\eta_B$ at the break in the derivative function $\phi_P$ (eq.3). Lower panel: Scaled PCS data for salol measured isothermally at high pressure [33].

**Fig.8** Upper panel: Dielectric relaxation times for salol ( experimental data for 0.1 MPa, calculated curves for P = 0.3 and 1 GPa, and V=0.7896 ml/g). Lower panel: Stickel functions, with linear fits over the respective ranges -4.58 < log(τ[s]) < 1.6 and -9.59 < log(τ[s]) < -7.44.

**Fig.9** Isobaric dielectric relaxation times for salol, PC, BMMPC, PDE, KDE and PCB62 versus $T_\alpha/T$ where $\tau(T_\alpha)$= 10s. Isochors were calculated at the volume at which τ=10s at atmospheric pressure; V=0.7907 (salol), 0.7558 (PC), 0.9067 (BMMPC), 0.7297 (PDE), 0.7748 (KDE) and V=0.6131 (PCB62) ml/g.

**Fig.10.** Isobaric steepness index at, $m_P$, versus pressure, calculated from the curves in Fig.9. The lines are to guide the eyes.

**Fig.11.** Log(τ) versus *P* for 1,2-polybutadiene measured at four temperatures. Open symbol are experimental data, solid symbols were calculated from data at atmospheric pressure using $\log(\tau) = \Im(TV^{1.9})$. Solid lines represent fits to eq.(5). The insert shows the $m_P$ calculated from the isotherms using $\log(\tau) \propto TV^{1.9}$.

**Fig.12.** Log(τ) versus $T^{-1}V^{-\gamma}$ for the six materials studied herein, along with the best fit of eq.(10). The fit parameters are given in Table 2. The residuals are plotted in the insets.

**Table captions**

**Table 1** Results from analysis of dielectric relaxation data for several materials. The $T_B$ were determined from the intersection of the two linear fits, and $\tau_B$ is the corresponding value of relaxation time.

**Table 2** Best fit parameters of the dielectric relaxation times to eq.(10)

| Material | | range of T [K] | $T_B$ [K] | $\log(\tau_B$ [s]) |
|---|---|---|---|---|
| *PDE* | Isobar 0.1 MPa | 295.0 - 415.1 | 321.7±9 | -3.9±1.5 |
| *PDE* | Isobar 100 MPa | 320.5 - 460.8 | 349.7±9 | -3.6±0.8 |
| *PDE* | Isobar 400 MPa | 384.6- 561.8 | 421.8±11 | -3.8±1 |
| *PDE* | Isochoric 0.7285 ml/g | 295.0 - 589.9 | 345±10 | -3.7±0.8 |
| *PCB62* | Isobar 0.1 MPa | 268.1-377.4 | 315.5±8 | -5.9±1 |
| *PCB62* | Isobar 200 MPa | 323.5-498.5 | 393.7±6 | -5.9±0.9 |
| *PCB62* | Isobar 600 MPa | 381.7-662.2 | 480.7±10 | -5.9±1 |
| *PCB62* | Isochoric 0.6131 ml/g | 259.5-690.1 | 393±10 | -5.8±0.9 |
| *PC* | Isobar 0.1 MPa | 159.1-276.0 | 188.6±7 | -7±0.9 |
| *PC* | Isobar 300 MPa | 183.7-336.9 | 221.3±9 | -6.9±.6 |
| *PC* | Isobar 600 MPa | 204.5-383.4 | 247.5±10 | -7.0±1 |
| *PC* | Isobar 1 GPa | 228.3-433.5 | 277.7±11 | -6.9±0.7 |
| *PC* | Isochoric 0.7562 ml/g | 159.1-382.4 | 202.9±8 | -7.2±0.7 |
| *KDE* | Isobar 0.1 MPa | 307.7-478.9 | 373.9±10 | -6.4±0.8 |
| *KDE* | Isobar 400 MPa | 386.1-678.4 | 506.9±12 | -6.4±0.6 |
| *KDE* | Isobar 1 GPa | 523.0-880.0 | 653.8±16 | -6.4±0.6 |
| *KDE* | Isochoric 0.7562 ml/g | 307.7-827.8 | 452.7±13 | -6.3±0.4 |
| *BMMPC* | Isobar 0.1 MPa | 259.1-359.7 | 315.4±10 | -6.2±0.9 |
| *BMMPC* | Isobar 200 MPa | 306.4-442.1 | 377±19 | -6.0±1 |
| *BMMPC* | Isobar 600 MPa | 383.4-565.9 | 480±22 | -6.1±1 |
| *BMMPC* | Isochoric 0.9032 ml/g | 259.2-613.5 | 422±40 | -6.2±1 |
| *Salol* | Isobar 0.1MPa | 219.3-315.4 | 253.2±5 | -6.3±0.7 |
| *Salol* | Isobar 300 MPa | 268.2-409.5 | 315.4±6 | -6.3±0.7 |
| *Salol* | Isobar 600 MPa | 308.8-491.2 | 372.1±7 | -6.6±0.5 |
| *Salol* | Isochoric 0.9032 ml/g | 219.1-458.1 | 290.2±5 | -6.3±0.3 |

**Table 1** Casalini and Roland

| Material | γ | A | B [Kml$^γ$g$^{-γ}$] | C [Kml$^γ$g$^{-γ}$] | Y$_0^{-1}$ [Kml$^γ$g$^{-γ}$] |
|---|---|---|---|---|---|
| *PDE* | 4.5 | -9.90±0.04 | 154.2±5.8 | 6.27±0.09 | 81.1±0.6 |
| *PC* | 3.7 | -11.161±0.008 | 150.4±1.4 | 3.42±0.05 | 57.69±0.19 |
| *KDE* | 4.5 | -10.4±0.01 | 291.6±4 | 17.95±0.1 | 122.6±0.5 |
| *BMMPC* | 8.5 | -10.2±1.4 | 393±294 | 44±27 | 170.3±12.8 |
| *salol* | 5.2 | -10.74±0.01 | 125.6±2.5 | 7.21±0.07 | 82.0±0.4 |
| PCB62 | 8.5 | -9.49±0.03 | 5.5±0.3 | 0.79±0.03 | 7.19±0.03 |

**Table 2** Casalini and Roland

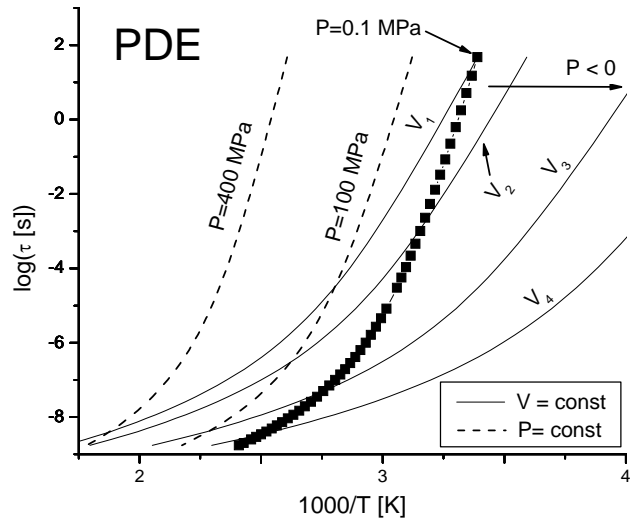

Fig.1: Casalini and Roland

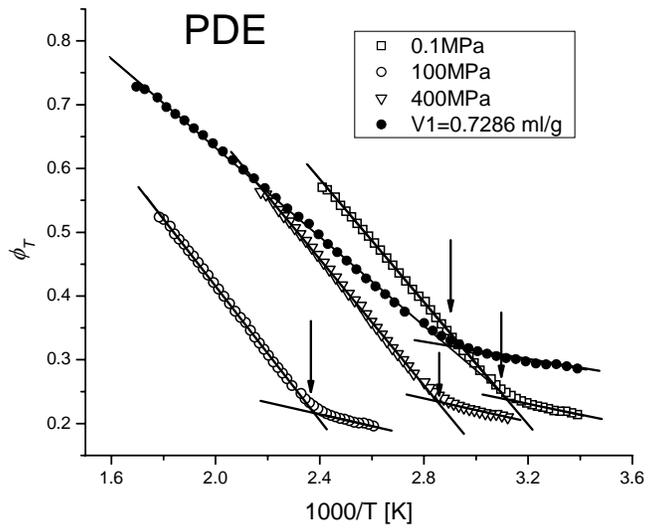

Fig.2 Casalini and Roland

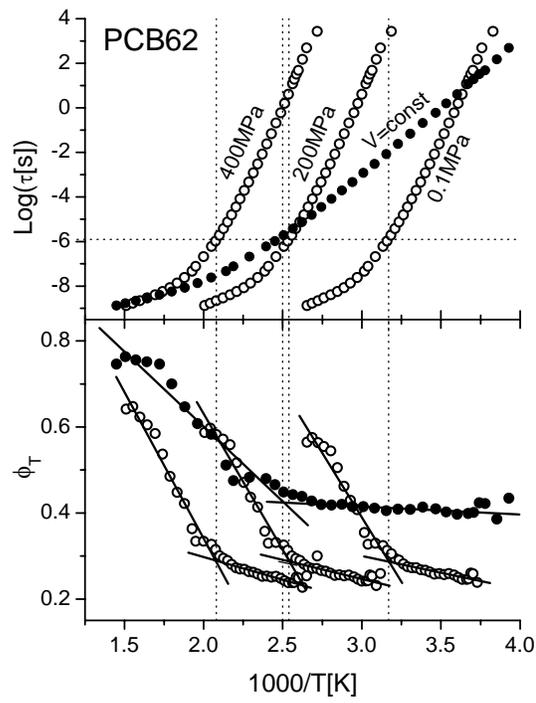

Fig.3 Casalini and Roland

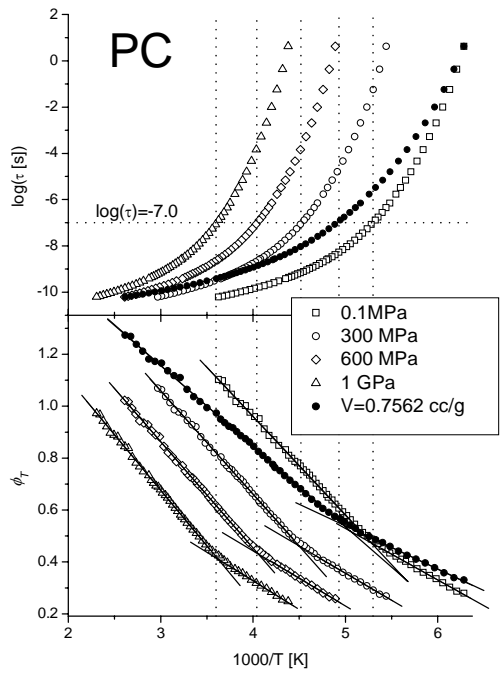

Fig.4 Casalini and Roland

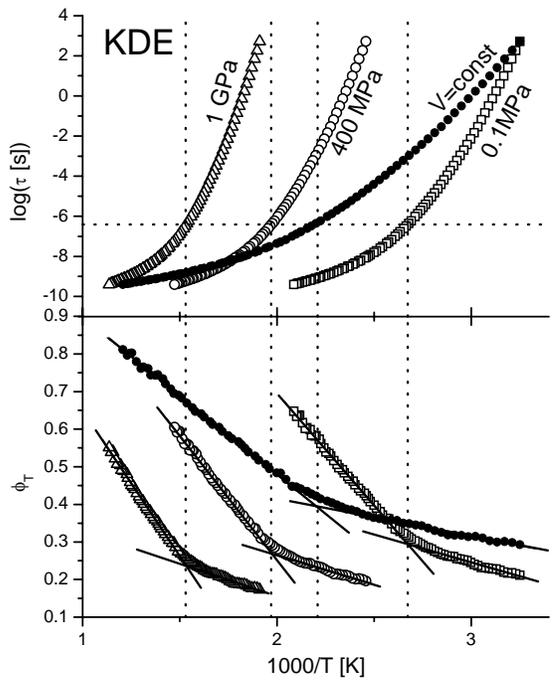

Fig.5 Casalini and Roland

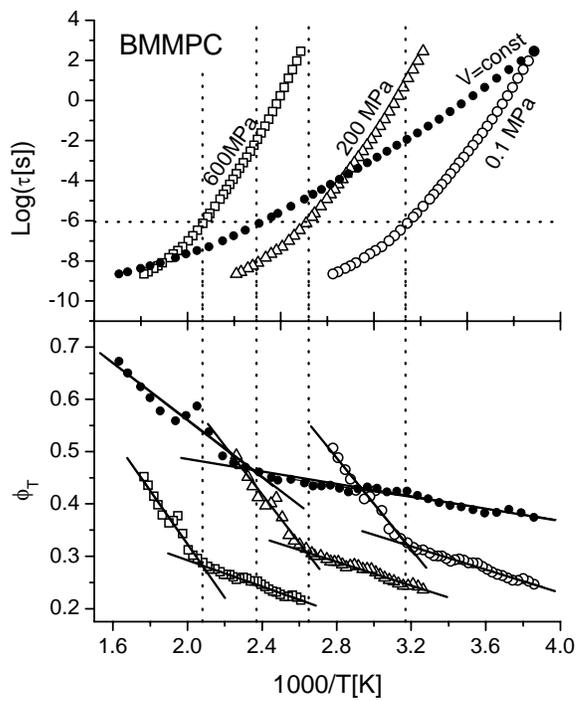

Fig.6 Casalini and Roland

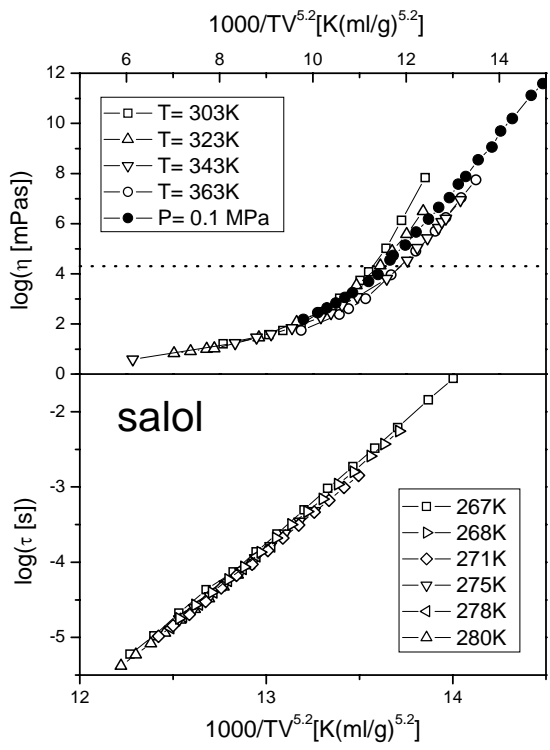

Fig.7 Casalini and Roland

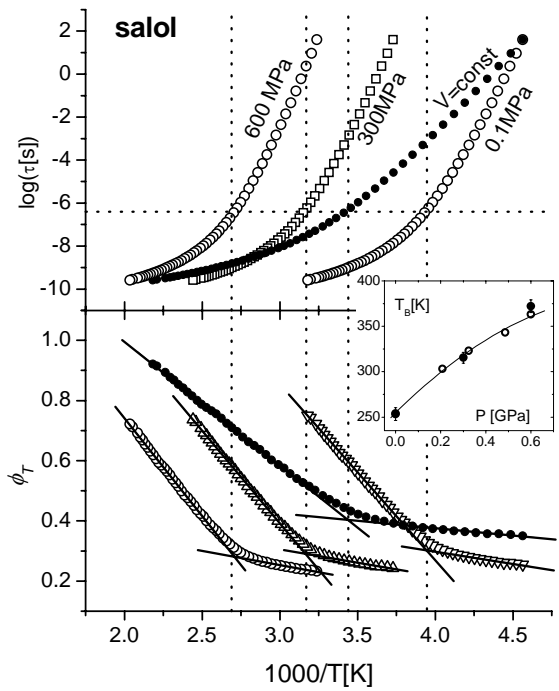

Fig.8 Casalini and Roland

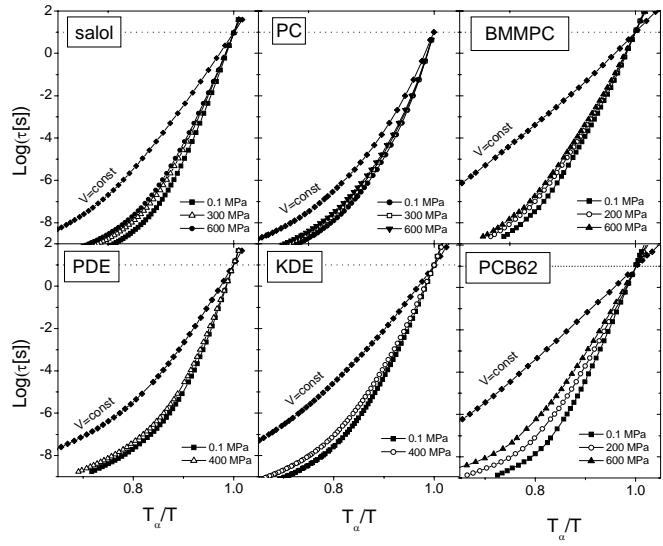

**Fig.9** Casalini and Roland

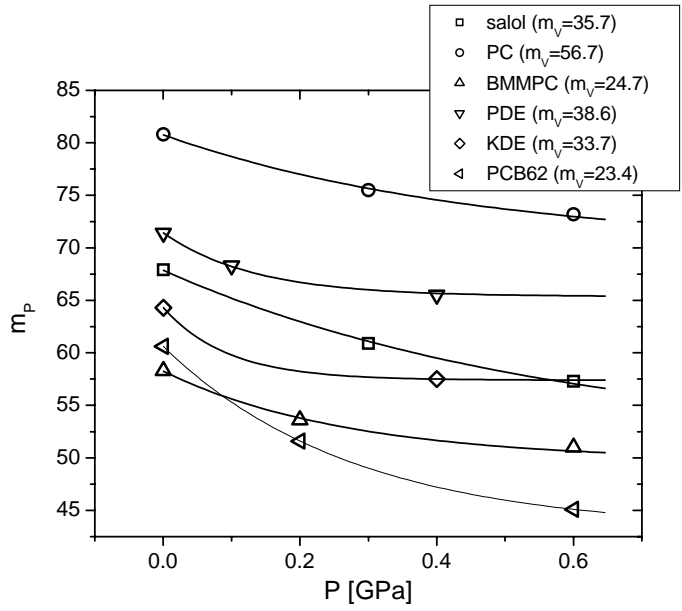

**Fig.10.** Casalini and Roland

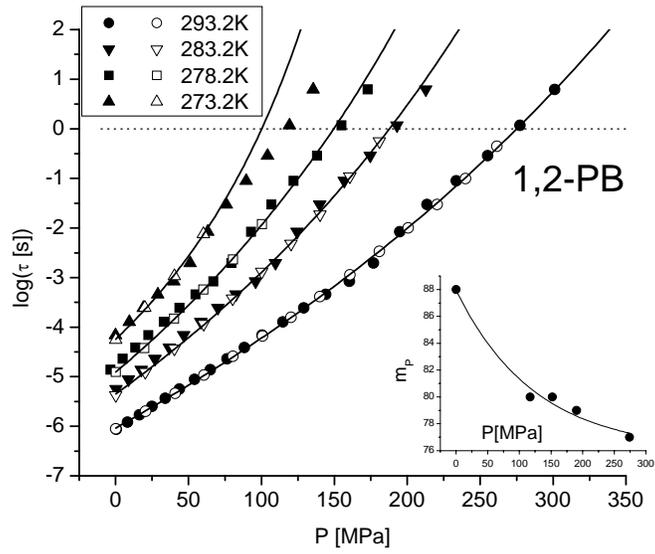

**Figure 11.** Casalini and Roland

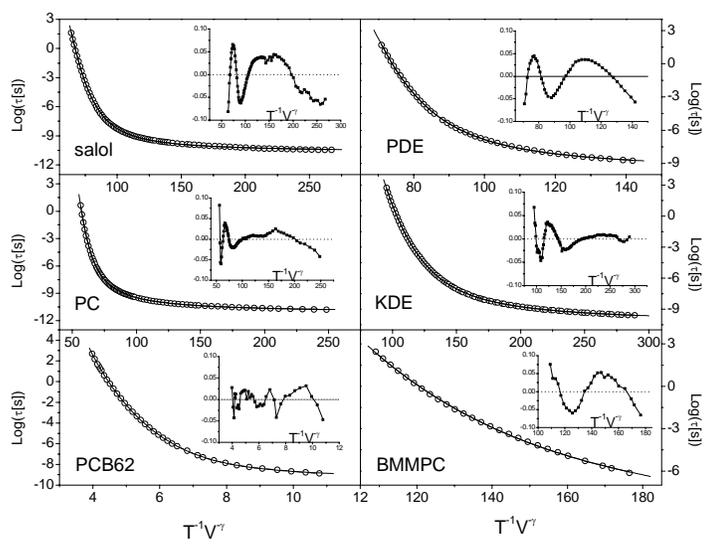

**Fig.12** Casalini and Roland